# Tuning near-gap electronic structure, interface charge transfer and visible light response of hybrid doped graphene and $Ag_3PO_4$ composite: Dopant effects

Chao-Ni He[1], Wei-Qing Huang[1*], Liang Xu[2,1], Yin-Cai Yang[1], Bing-Xin Zhou[1], Gui-Fang Huang[1#], P. Peng[2], Wu-Ming Liu[3§]

[1] Department of Applied Physics, School of Physics and Electronics, Hunan University, Changsha 410082, China

[2] School of Materials Science and Engineering, Hunan University, Changsha 410082, China

[3] Beijing National Laboratory for Condensed Matter Physics, Institute of Physics, Chinese Academy of Sciences, Beijing 100190, China

---

[*]. Corresponding author. *E-mail address:* wqhuang@hnu.edu.cn
[#]. Corresponding author. *E-mail address:* gfhuang@hnu.edu.cn
[§]. Corresponding author. *E-mail address:* wliu@iphy.ac.cn

**Abstract:** The enhanced photocatalytic performance of doped graphene(GR)/semiconductor nanocomposites have recently been widely observed, but an understanding of the underlying mechanisms behind it is still out of reach. As a model system to study the effect of dopants, we investigate the electronic structures and optical properites of doped GR/$Ag_3PO_4$ nanocomposites using the first-principles calculations, demonstrating that the band gap, near-gap electronic structure and interface charge transfer of the doped GR/$Ag_3PO_4$(100) composite can be tuned by the dopants. Interestingly, the doping atom and C atoms bonded to dopant become active sites for photocatalysis because they are positively or negatively charged due to the charge redistribution caused by interaction. The dopants can enhance the visible light absorption and photoinduced electrons transfer. We propose that the N atom may be most appropriate doping for the GR/$Ag_3PO_4$ photocatalyst. This work can rationalize the available experimental results about N-doped GR-semiconductor composites, and enriches our understanding on the effect of dopants in the doped GR-based composites for developing high-performance photocatalysts.

Semiconductor photocatalysis is a promising technology to address problems in environmental remediation and energy utiliztion, such as water splitting for hydrogen production.[1-4] Photocatalytic semiconductors are generally metal oxides, nitrides, or sulfides.[1] Among them, titanium dioxide ($TiO_2$) has proven to be one of the most suitable photocatalysts due to its strong oxidation ability, chemical and biological inertness, and low cost.[5,6] However, the wide band gap (∼3.2 eV) implies that its photocatalytic activity is only triggered under ultraviolet light (UV), leaving about 95% solar energy useless.[7-9] Developing novel photocatalysts with outstanding performance under visible (vis) light is therefore pursued.

Photoexcited carriers occurred in photocatalysis are involved in two competing processes: (i) diffusion to the surface such that chemical reactions with adsorbed molecules can occur; and (ii) recombination, which decreases the number of active carriers consumed by chemical reactions on the surface.[10] The rapid recombination rate of photogenerated electron-hole pairs within semiconductors (such as $TiO_2$) results in the low quantum efficiency, thus limiting their practical application. The most effective strategy is to couple semiconductors with other materials forming heterojunction, in which the electrons and holes are separated via interfacial charge transfer. As the paradigm in this respect, graphene (GR)-based semiconductor nanocomposite has recently gained increasing interest.[11,12] The high specific surface area and superior electron mobility of GR are in favor of the interfacial charge transfer between GR and semiconductors, thus hindering the recombination process of electron-hole pairs in photocatalysis and enhancing the photocatalytic activity of composites. Interestingly, experimental studies have demonstrated that the photocatalytic performance of GR-semiconductor composites can be further improved by doping.[13-26] Jia et al. first reported that N-doped GR (N-GR)/CdS nanocomposites have a higher photocatalytic activity than pure CdS, and the cocatalyst of N-GR as a protective layer can prevent CdS from photocorrosion under light irradiation.[13] Subsequently, the enhanced photocatalytic activities of doped GR-semiconductor composites, such as N-GR/ZnSe,[14] g-$C_3N_4$/N-GR/$MoS_2$,[15] N-GO/$MoS_2$,[16] N-GR/AgX@Ag (X= Br, Cl),[17] N-GR/$Fe_2O_3$,[18] N-GR/Pd@PdO,[19] B-GR/$TiO_2$,[20,23] N-GR/ZnO,[21] N-GR/$TiO_2$,[22,24,26] N-GR/ZnS,[25] have also been verified experimentally. It is generally assumed that the improved activities can be ascribed to the synergistic effects of more light harvest, enhanced adsorption capacity and more efficient separation of photogenerated electron-hole pairs after integrated with doped GR. However, the underlying mechanisms of the dopant effect on the charge transfer and photocatalytic performance of the composites have not yet been revealed.

In this work, the effects of dopants is systematically explored via a model doped GR/$Ag_3PO_4$ composite using density functional theory (DFT) calculations. The choice of $Ag_3PO_4$ is motivated by its important potential applications as a vis-light-sensitive photocatalyst, due to its extremely high photo-oxidative capabilities for $O_2$ evolution from water and for the decolorization of organic dyes, its quantum efficiencies up to nearly 90% under vis-light irradiation.[27-31] However, $Ag_3PO_4$ crystal is slightly soluble in aqueous solution, which greatly reduces its structural stability. Moreover, during the photocatalytic process, $Ag^+$ is usually transformed into Ag, resulting in the photocorrosion of $Ag_3PO_4$ due to the absence of electron acceptors.[27] The instability of $Ag_3PO_4$ photocatalytic limits its practical application as a recyclable highly efficient photocatalyst. Coupling semiconductors with $Ag_3PO_4$ is recognized to be a viable strategy to enhance its photocatalytic activity and stability.[8,32-43] In particular, the enhanced photocatalytic performance of GR/$Ag_3PO_4$ composite has been investigated both experimentally and theoretically.[39, 44] Therefore, the doped GR/$Ag_3PO_4$ composite is expected to be prime candidate to uncover the dopant effect for its significance not only for use under vis-light, but also for insightful studies of photocatalytic mechanisms. Here, four dopants (B, N, S, and P) are chosen as representative to uncover their role in the doped GR/$Ag_3PO_4$ composites. The impurity states located at near the gap will change the electronic structure, thus tuning the band gap and vis-light response of the doped GR/$Ag_3PO_4$ composites. Interestingly, the doping atom and the C atoms bonded to dopant become positively or negatively charged, acting as active sites for photocatalysis. The $N_C$-GR/$Ag_3PO_4$(100) composite is a promising photocatalyst due to its highest visible light absorption and stability. These results provide the fundamental understanding of the effect of dopants, which is important in the future design of vis-light-harvesting doped-GR-based composites.

## Results and Discussions

**Structural properties and formation energy.** The representative views of GR/$Ag_3PO_4$(100) composites are shown in Fig. 1. Parts (a) and (b) present the top and side views of the GR/$Ag_3PO_4$(100) interface used in our calculations, respectively. The blue atom is the center C atom of GR sheet, which is replaced by B, N, S, and P in the $B_C$-GR/$Ag_3PO_4$(100), $N_C$-GR/$Ag_3PO_4$(100), $S_C$-GR/$Ag_3PO_4$(100), and, $P_C$+$V_C$-GR/$Ag_3PO_4$(100) composites, respectively. Part (c) of Fig. 1 is given the top view of the $P_C$+$V_C$-GR/$Ag_3PO_4$(100) composite, while part (d) for side view of the $S_C$-GR/$Ag_3PO_4$(100) composite after optimization. Geometry optimization has first been performed for these systems using the conjugate gradient method. The distances from the $Ag_3PO_4$(100) surface to GR/doped GR sheets after optimization for the GR/$Ag_3PO_4$(100) and doped GR/$Ag_3PO_4$(100) composites are listed in Table 1. For the pure GR/$Ag_3PO_4$(100) composite, the farthest and closest distances between the GR sheet and the top of the $Ag_3PO_4$(110) surface is calculated to be 2.65 and 2.61 Å, respectively, which is about equal to those between the GR sheet and other materials (2.65 Å for $TiO_2$(110)/GR,[46] 2.422− 2.866 Å for ZnO(0001)/GR[47]). This indicates that the GR sheet is slightly distorted because the forces on different C atoms, caused by the interfacial interaction, are vary, owing to the different arrangement of atoms at the top layer of $Ag_3PO_4$(100) surface.

The effect of dopants on the interface distance of doped GR/$Ag_3PO_4$(100) composites can be clearly seen from Table 1. The variation of interface distance of doped GR/$Ag_3PO_4$(100) composites depends on the impurity atom. The B dopant on the carbon lattice site induces a small increase (about 0.02~0.03 Å) of interface distance between the B-dope GR sheet and $Ag_3PO_4$(100) surface. The same increase of interface distance also happens in the $P_C+V_C$-GR/$Ag_3PO_4$(100) composite. Surprisingly, the interface distance between the N-dope GR sheet and $Ag_3PO_4$(100) surface is significantly increased (~ 0.5 Å) compared to that in pure GR/$Ag_3PO_4$(100) composite. Although the interface distances are changed in these composites, the doped GR sheets are still quite flat, indicating that the interfacial interactions are indeed vdW rather than covalent, in accordance with previous studies.[44] On the contrary, the S dopant leads to a local distinct deformation of the GR sheet, as shown in Fig. 1(d). This is because the bond length of S-C (1.74 Å) is longer than that of C-C (1.42 Å), while those of B-C and N-C are 1.48 and 1.41 Å, respectively, very close to 1.42 Å. Thus, the farthest and closest distances between the S-doped GR sheet and the top of the $Ag_3PO_4$(110) surface are 2.71 and 1.73 Å, respectively. The microscopic crumpling around the S dopant can be comparable in size to the suspended graphene sheets, which exhibit intrinsic microscopic roughening such that the surface normal varies by several degrees and out-of-plane deformations reach 1 nm.[45] The change of interface distance indicates the variations of interface interaction between doped GR and $Ag_3PO_4$, which will result into the change of electron transfer in these composites. Meanwhile, the atoms in the top two layers of $Ag_3PO_4(100)$ surface have been rearranged, implying that the electron transfer occurs at the interface, as will be shown below.

To assess the effect of dopants on the relative stability of these doped composites, the interface binding energies, $\Delta E_f$, of these composites are estimated according to the following equation:

$$\Delta E_f = E_{comb} - E_{X-GR} - E_{Ag_3PO_4(100)} \qquad (1)$$

where $E_{comb}$, $E_{X-GR}$, and $E_{Ag_3PO_4(100)}$ represent the total energy of the relaxed X-GR/$Ag_3PO_4(100)$ (X=B, N, S, P+$V_C$), X-GR (X=B, N, S, P+$V_C$) sheet, and pure $Ag_3PO_4(100)$ surface, respectively. By this definition, the negative $\Delta E_f$ indicates that the composite is stable. The lower $\Delta E_f$ will make the crystal to approach to the lower energy state and therefore be steadier, sugggesting that the composites can be experimentally synthesized more easily. The $\Delta E_f$ are calculated to be −0.77, −5.12, -5.48, and -7.34 eV of the GR/$Ag_3PO_4(100)$, $B_C$-GR/$Ag_3PO_4(100)$, $N_C$-GR/$Ag_3PO_4(100)$, and $P_C$+$V_C$-GR/$Ag_3PO_4(100)$ composites, as listed Table 1. The $S_C$-GR/$Ag_3PO_4(100)$ composite has the highest $\Delta E_f$ (2.220 eV), due to its biggest deformation. On the contrary, the $\Delta E_f$ of other three doped composites are more negative than that of pure GR/$Ag_3PO_4(100)$ composite. Thus, the $B_C$, $N_C$ and $P_C$+$V_C$ doped GR sheets are more easily to couple with $Ag_3PO_4$ or other semiconductors to form composites. These theoretical predictions are in line with the available experimental results, such as N-GR/ZnSe[14]($Fe_2O_3$,[18] ZnO,[21] $TiO_2$,[22, 24, 26] ZnS,[25] AgX@Ag (X= Br, Cl)[17]) and B-GR/$TiO_2$.[20,22]

**Density of States.** Incorporating impurity atom into semiconductors to engineer their electronic structures and photocatalytic performances have widely been used as an effective strategy. The influence of dopants on the properties of the composites containing two-dimensional layered material has not been explored theoretically so far. To evaluate the effect of dopants, it is insightful to analyze the electronic structure of doped-GR-based composites. The density of states (DOSs) of five models are displayed in Fig. 2. As its (100) surface is exposed, $Ag_3PO_4$ becomes a direct band gap semiconductor and the band gap, $E_g$, decreases to 2.15 eV from 2.45 eV.[44] The GR/Ag3PO4(100) has an $E_g$ of 0.53 eV due to the Fermi level down-shifting by 0.655 eV with respect to the Dirac point of GR, which extends its absorption spectrum covering the entire visible region, even infrared light.[44] Fig. 2 shows that, for pure $GR/Ag_3PO_4(100)$ composite, the upper part of valence band (VB) consists of the states of GR sheet, whereas the bottom of conduction band (CB) is composed by orbitals from the $Ag_3PO_4(100)$ surface. Under light irradiation, electrons in the top VB can be directly excited to the CB of $GR/Ag_3PO_4(100)$, producing well-separated electron-hole pairs. The most important feature in Fig. 2 is the band alignment between the doped GR sheet and the $Ag_3PO_4(100)$ surface. As shown in the left column of Fig. 2, the main shapes of the calculated DOSs projected on two different components in the doped $GR/Ag_3PO_4(100)$ composites are similar to those of the TDOSs of isolated doped GR sheets and the $Ag_3PO_4(100)$ surface, respectively (note that the DOSs of doped GR sheets are not given here). This is due to larger separation space between them (about 2.6−3.2 Å), suggesting that the doped GR-$Ag_3PO_4(100)$ surface interaction is weak due to the absence of covalent bonding upon formation between the interfaces. The interaction between the doped GR sheets and $Ag_3PO_4(100)$ surface varies with different dopants. From Fig. 2, the VB offset (VBO) and the CB offset (CBO) between GR sheets and $Ag_3PO_4(100)$ surface appear in these composites. This kind of the band alignment between the two components demonstrate that a type-II heterojunction is formed, which is key to enhance the photoactivity of the doped $GR/Ag_3PO_4(100)$ composites.

Doping can tune the band gap of doped GR/Ag$_3$PO$_4$(100) composites, as listed in Table 2. The B or S substituting on the C site in the GR sheet will decrease the band gap of the composites. Similarly, the P$_C$+V$_C$-GR/Ag$_3$PO$_4$(100) composite has a smaller band gap (0.12 eV), compared to pure GR/Ag$_3$PO$_4$(100). The decrease of band gap, due to B, S, and P dopants, stems from the negative movement of the CB bottom, relative to the Fermi level (Figs. 2 (b1, d1, e1)). In contrast, N atom doped into the GR sheet as impurity increases the band gap from 0.53 to 1.44 eV of N$_C$-GR/Ag$_3$PO$_4$(100) composite. The origin of band gap broaden can be traced by the movement of the VB maximum (VBM) and the CB minimum (CBM). Compared to the Fermi level, there is a significantly raise of the CBM compared to the pure GR/Ag$_3$PO$_4$(100) composite. Note that the CBM consists mainly of O 2p states from the top layer of the Ag$_3$PO$_4$(100) surface. A nitrogen dopant atom has three unpaired electrons with strong N 2p character, similar to its neighboring O 2p states, leading to the hybridization between N and its neighboring O atoms. Figs. 2 (c2) and (d2) show that the N 2p orbital above the Fermi level is much higher than those of S dopant. Since the N 2p orbital energy is higher than the O 2p orbital energy, the overlap of N 2p and O 2p states at the bottom of CB then elevates the CBM, as well as the donor level to be above the VB. Therefore, doping with N impurity into GR sheet widens the band gap of N$_C$-GR/Ag$_3$PO$_4$(100) composite, whereas the S dopant reduces the band gap of S$_C$-GR/Ag$_3$PO$_4$(100) composite. The N dopant can also broaden the band gap of Cu$_2$O.[48] The cases of B$_C$-GR/Ag$_3$PO$_4$(100) and P$_C$+V$_C$-GR/Ag$_3$PO$_4$(100) composites are quite different: the B 2p and P 3p states are located at the Fermi level, i. e., to modify the VB edge (p-type doping) (see Figs. 2 (b2) and (e2)), thus the decrease of band gaps resulting in an upward shift of the VBM (see Figs. 2 (b1) and (e1)). These results indicate that the substitutional doping of GR sheet not only tune the band gap, but also change the ingredients of near-gap electronic structure of the doped-GR-based composites, which will influence the light absorption.

**Mulliken Population, Charge Transfer and Mechanism Analysis.** The interactions between GR and $Ag_3PO_4(100)$ surface would induce the charge distribution fluctuations of GR sheet.[44] To quantitatively analyze the charge variation caused by doping in the doped $GR/Ag_3PO_4(100)$ composites, the Mulliken population analysis of the plane-wave pseudopotential calculations has been performed for five composites. Figs. 3 and 4 render the results of the Mulliken charges on the atoms near the dopants, in which selected typical values are denoted. Although the C atom in the pure GR has a Mulliken charge of zero electrons, those C atoms in the $GR/Ag_3PO_4(100)$ composite have different Mulliken charges because the arrangement of atoms under various C atoms is different (Fig. 3(a)). The redistribution of charge of doped GR sheet mainly depends on the incorporation of impurity atoms into carbon lattice. As shown in Figs. 3 (b-e), the doped B, N, S, and P atoms have a Mulliken charge of +0.51, −0.23, +0.84, and +1.06, respectively. Compared Fig. 3 (a) with Figs. 3 (b-e), one can see that the C atoms bonded to the dopants have much larger Mulliken charges than those in the pure $GR/Ag_3PO_4(100)$ composite. Interestingly, the three C atoms bonded to N atom have positive Mulliken charges (+0.14, +0.15), whereas those bonded to B (S, P) atom have negative Mulliken charges. This can be ascribed to the different atomic radiuses, electronegativities, and their bond length with C atom, of these dopants. Moreover, some C atoms become positively charged (not given in Fig. 3), while others are negative in these composites. Those atoms, especially the dopants, with positive effective charge will facilitate the adsorption of some species with negative charges, thus becoming active sites. The dopants in GR sheets also change the charge redistribution of the $Ag_3PO_4(100)$ surface, which are shown in Fig. 4. In particular, there are obvious changes of the Mulliken charges of Ag atoms under the dopants. For example, Fig. 4(c) shows that the Mulliken charges of the Ag atoms in the first, second and third layers are reduced from 0.85 to 0.73, 0.56 to 0.51, 0.57 to 0.55, respectively, compared to those in undoped GR/Ag$_3$PO$_4$(100) composite (Fig. 4(a)). The value variation is decreased gradually is due to the elongation of the distance between the dopants and the Ag atoms in different layers. However, the electrons of O atoms in the top layer are almost no change. Therefore, incorporating substitutional impurity into the GR sheet is likely to be one effective strategy to improve stability of GR/Ag$_3$PO$_4$(100) composites.

The rearrangements of atoms near the interface and the strong change of DOSs imply a substantial charge transfer between the involved constituents. This can be visualized (as shown in Figs. 5 and 6) by the three-dimensional charge density difference $\Delta\rho=\rho_{X-GR/Ag_3PO_4(100)}-\rho_{Ag_3PO_4(100)}-\rho_{X-GR}$ (X=B, N, S, P+V$_C$), where $\rho_{X-GR/Ag_3PO_4(100)}$, $\rho_{Ag_3PO_4(100)}$, and $\rho_{X-GR}$ are the charge densities of the composite, Ag$_3$PO$_4$(100) surface, and free-standing X-GR in the same configuration, respectively. Similar to the case of undoped GR/Ag$_3$PO$_4$(100) composite, the strong charge accumulations are found just above the Ag atoms in the top layer, whereas the regions of charge depletion appear both on the lower side of the doped GR (facing the surface) just above the O atoms in the top layer and on the higher side of the Ag atoms in the top layer in the ground electronic state. It can be clearly seen from Figs. 5 and 6 that the effects of dopants on the interface charge transfer are apparent. The relative order of the amount of charge transfer for the doped composites is found to be N$_C$-GR/Ag$_3$PO$_4$(100) > S$_C$-GR/Ag$_3$PO$_4$(100) > P$_C$+V$_C$-GR/Ag$_3$PO$_4$(100) > B$_C$-GR/Ag$_3$PO$_4$(100). To offer quantitative results of charge redistribution, parts (c) of Figs. 5 and 6 plot the planar averaged charge density difference along the direction perpendicular to the Ag$_3$PO$_4$(100) surface. For the B$_C$-, N$_C$-, S$_C$-, and P$_C$+V$_C$-GR/Ag$_3$PO$_4$(100) composites, the largest efficient electron accumulations localized above the Ag atoms in the top layer are respectively about $16.0 \times 10^{-4}$, $30.0 \times 10^{-4}$, $20.0 \times 10^{-4}$, and $18.0 \times 10^{-4}$ e/Å$^3$, while the largest efficient electron depletions on the lower side of the GR are respectively about $-18.0 \times 10^{-4}$, $-48.0 \times 10^{-4}$, $-31.0 \times 10^{-4}$, and $-18.0 \times 10^{-4}$ e/Å$^3$. These values are larger than those in the undoped GR/Ag$_3$PO$_4$(100) composite, indicating that the interaction between doped-GR sheet and Ag$_3$PO$_4$(100) surface becomes stronger due to doping. This can be further verified by the efficient electron accumulation above the doped GR sheets, which is more than that in the undoped GR/Ag$_3$PO$_4$(100) composite. A further charge analysis based on the Bader method can give the quantitative results of charge transfer between the two constituents, which are listed in Table 2. One can see that 0.39 electrons transfers from GR sheet to Ag$_3$PO$_4$(100), in agreement with the previous results.[44] However, the amount of charge transferred at the B$_C$-GR/Ag$_3$PO$_4$(100) interface is 0.15 electron from B$_C$-doped GR sheet to Ag$_3$PO$_4$, which is less than that at the GR/Ag$_3$PO$_4$(100) interface. The decreased amount of charge transferred is because of the longer average equilibrium distance (3.08-3.11 Å) between the B$_C$-doped GR sheet between Ag$_3$PO$_4$(100) surface and the doped B atom with smaller electronegativity than C atom, leading to that some electron transfers from B atom to C atom in the GR sheet. On the contrary, the amounts of charge transferred in the other three doped GR/Ag$_3$PO$_4$(100) composite are larger than that in undoped GR/Ag$_3$PO$_4$(100) one, which may be related to their electronegativities and structural deformation of doped GR sheets.

The underlying mechanism for different interface charge redistributions can be traced to the electrostatic potential distribution in the doped GR/Ag$_3$PO$_4$(100) composites. The profile of the planar averaged self-consistent electrostatic potential for these composite as a function of position in the -direction is displayed in Figs. 5, 6 (a) and (d). Similar to the case of undoped GR/Ag$_3$PO$_4$(100) composite, the periodic lattice potential in the Ag$_3$PO$_4$(100) lattice is clear although it has some distortion due to the atoms in the upper layers having a slight movement compared to their positions in bulk $Ag_3PO_4$. The higher potential near the doped GR sheets leads to a potential well formed at interfaces. The appearance of such a large built-in potential well can effectively hinder the recombination of photogenerated charge carriers in the doped GR/$Ag_3PO_4$(100) composite. During the photocatalytic process, photogenerated electrons migrated to the $Ag_3PO_4$(100) surface could be pumped to the doped GR sheets, resulting in the net efficient electrons accumulation at doped GR sheets, because the built-in potential is large enough to drive efficient charge separation in the composites. Among these dopants, substitutional N doping can push the plane of largest potential at the GR sheet to the $Ag_3PO_4$(100) surface, as shown in Fig. 5(d). The closer the largest potential gets to the $Ag_3PO_4$(100) surface, the higher the migration efficiency of electrons at the $Ag_3PO_4$(100) surface. These results partly offer a physical interpretation for the enhanced photocatslytic activities of the N-doped GR/semiconductor composites synthesized by experiments.[13-18, 21, 22, 24-26, 49, 50]

**Optical Properties.** To investigate the influence of dopants on the optical properties, we have calculated the absorption spectra of the undoped and doped GR/$Ag_3PO_4$(100)composites, as shown in Fig. 7. One can see, the undoped GR/$Ag_3PO_4$(100) composite has a significant absorption at a wide region from 300 to 800 nm. The remarkable feature of its absorption curve is three resonant-like peaks at about 300, 440, and 780 nm, which has been discussed previously.[44] As the impurity atom is doped into the GR sheet in the composites, the absorption curves have more than one resonant-like peak (3 peaks for $B_C$- and $P_C+V_C$-GR/$Ag_3PO_4$(100), 4 peaks for $N_C$- and $S_C$-GR/$Ag_3PO_4$(100) composites), due to the transitions between different energy levels. Compared to the undoped GR/$Ag_3PO_4$(100) composite, the weaker absorption intensity of the doped GR/Ag$_3$PO$_4$(100) composite at long wavelength range ($\lambda$>600 nm) can be attributed to the change of the near-gap electronic structures caused by the dopants. Fortunately, the doped GR/Ag$_3$PO$_4$(100) composites have strong absorption in the UV and vis-light range (ranging from 300 to 600 nm) which is more importantly for the photocatalysis. In particular, the higher resonant-like peaks at about 400~440 nm, due to doping, are especially beneficial to enhance the photocatalytic performance of the doped GR/Ag$_3$PO$_4$(100) composites, according to the fact that the light with wavelength of ~440 nm might be the most appropriate visible light for generation of radical species in the GR/Ag$_3$PO$_4$(100) composite.[39,44] Fig. 7 illuminates that the N$_C$-GR/Ag$_3$PO$_4$(100) composite has excellent absorption, followed by P$_C$+V$_C$-GR/Ag$_3$PO$_4$(100) composite. Therefore, it can be concluded that doping with substitutional impurities into the GR sheet is an effective strategy to enhance the photocatalytic performance of the GR/Ag$_3$PO$_4$(100) composites.

We now propose, based on the above results, that the N atom may be most appropriate doping for the GR/Ag$_3$PO$_4$ photocatalyst. Firstly, the N$_C$-GR/Ag$_3$PO$_4$(100) composite can be easily prepared due to its lower formation energy (-5.48 eV). Secondly, the N dopant results in the more stronger absorption in the UV and vis-light range, especially from 350 to 600 nm, which is one of the important factors for a high-efficiency photocatalyst. Thirdly, the C atoms bonded to N dopant become positively charged while the N dopant is negative in the N$_C$-GR/Ag$_3$PO$_4$(100) composites. Those C atoms (the N dopant) with positive (negative) effective charge, as active sites, will facilitate the adsorption of some species with negative (positive) charges, thus enhancing the photocatalytic activity. Most importantly, the photocatalytic activity and stability of the GR/Ag$_3$PO$_4$(100) composite can be greatly improved by doping N atom in the GR sheet, owing to the fact that the highest potential plane will be pushed away from the GR sheet to the interface, thus promoting the migration of photogenerated electrons at the Ag$_3$PO$_4$ surface. Fortunately, this proposal here can be largely demonstrated by the available experimental results about N-doped GR-semiconductor composites.[13-18, 21, 22, 24-26, 49, 50]

## Summary

We have performed the first *ab initio* DFT calculations to explore the effects of dopants on the structural, electronic and optical properties, charge transfer at the interface formed between the B$_C$-, N$_C$-, S$_C$-, and P$_C$+V$_C$-GR sheets and a Ag$_3$PO$_4$(100) surface in the context of stable vis-light photocatalyst. The band gap, near-gap electronic structure and interface charge transfer of the doped GR/Ag$_3$PO$_4$(100) composite can be tuned by the dopants. Moreover, the substitutional doping in the GR sheet plays an essential role in the absorption of vis-light, in the promotion of electrons to the CB, and in photoinduced electrons transfer to reducible adsorbates of doped GR/Ag$_3$PO$_4$(100) composites. We propose that the N atom may be most appropriate doping for the GR/Ag$_3$PO$_4$ photocatalyst. This work provides the theoretical results to rationalize the available experimental results about N-doped GR-semiconductor composites, and enriches our understanding on the effect of dopants in the doped GR-based composites for developing high-performance photocatalysts.

## Methods

**Theoretical model.** We first construct a supercell (12.01×12.25×18 Å$^3$) of GR/Ag$_3$PO$_4$(100), a 5 × 6 single GR layer containing 60 carbon atoms sits on a 2×2 seven atomic layer stoichiometric Ag$_3$PO$_4$(100) surface (i.e., the most stable one among the low index surfaces) slab containing 64 atoms with three bottom layers fixed at bulk position. A 15 Å thick vacuum layer is above the GR sheet to avoid artificial interaction. The whole system contains totally 124 atoms with 736 valence electrons. To explore the dopant effect in the heterojunction, four substitutional dopants, B, C, S, and P, are chosen to substitute C atom of GR sheet. It should be pointed out that when one C atom is substituted by only a P atom, the deformation of doped GR sheet is very large, the vertical displacement of P atom more than 1.0 nm relative to the GR plane. This can be eliminated by removing one nearest neighbor C atom. Four doped GR/Ag$_3$PO$_4$(100) composites are: B$_C$-GR/Ag$_3$PO$_4$(100) (B atom substituting C atom of GR), N$_C$-GR/Ag$_3$PO$_4$(100) (N atom substituting C atom of GR), S$_C$-GR/Ag$_3$PO$_4$(100) (S atom substituting C atom of GR), and P$_C$+V$_C$ -GR/Ag$_3$PO$_4$(100) (P atom substituting C atom of GR and one neighbor C atom removed ). For comparison, undoped GR/Ag$_3$PO$_4$(100) composite are also studied.

**Electronic structure calculation.** The calculations are performed using the CASTEP code[51] based on first-principles DFT. The local density approximation (LDA) with inclusion of the van der Waals (vdW) interaction is chosen because long-range vdW interactions are expected to be significant in these complexes. However, LDA has been known generally to underestimate the energy gap of semiconductor, resulting into an overestimate for photoinduced electrons transfer in photocatalytic process. To correct this band gap problem, all of the theoretical calculations are performed using the DFT/LDA+U method implemented in the plane wave basis CASTEP code. We have performed extensive tests to determine the appropriate U parameters for Ag 4d states, which reproduced the correct energy gap (2.48 eV) for cubic Ag$_3$PO$_4$. The appropriate Hubbard U values for Ag 4d, O 2p, and P 3p are 7.2, 7.0, and 7.0 eV, respectively. The states of O: 2s$^2$ 2p$^4$, Ag: 4d$^{10}$ 5s$^1$, C: 2s$^2$ 2p$^2$, P: 3s$^2$ 3p$^4$, S: 3s$^2$ 3p$^4$, N: 2s$^2$ 2p$^3$, B: 2s$^2$ 2p$^1$, are treated as valence states. The plane wave cut off is set to 400 eV. The k mesh of 2×2×1 and 4×4×1 is respectively used for geometry optimizations and for calculating density of states using the Monkhorst-Pack scheme. In the geometrical optimization, the maximum ionic displacement is within 0.002 Å, and the total stress tensor is reduced to the order of 0.1 GPa. All geometry structures are fully relaxed until the convergence criteria of energy and force are less than $10^{-6}$ eV/atom and 0.01 eV/Å, respectively.

**Optical response.** The strong light absorption is one fundamental premise for a high-efficiency photocatalyst except for a low recombination rate of photogenerated charge carriers and suitable redox potentials. For the semiconductor material, the dielectric function is mainly connected with the electronic response. The imaginary part $\varepsilon_2$ of the dielectric function $\varepsilon$ is calculated from the momentum matrix elements between the occupied and unoccupied wave functions, given by[52]:

$$\varepsilon_2 = \frac{ve^2}{2\pi\hbar m^2 \omega^2} \int d^3k \sum_{n,n'} |\langle kn|p|kn'\rangle|^2 f(kn)(1 - f(kn'))\delta(E_{kn} - E_{kn'} - \hbar\omega) \quad (2)$$

Where $\hbar\omega$ is the energy of the incident photon, $p$ is the momentum operator $r(\hbar/i)(\partial/\partial x)$, $(|kn\rangle)$ is a crystal wave function and $f(kn)$ is Fermi function. The real part $\varepsilon_1$ of the dielectric function $\varepsilon$ is evaluated from the imaginary part $\varepsilon_2$ by Kramer–Kronig transformation. The absorption coefficient $I(\omega)$ can be derived from $\varepsilon_1$ and $\varepsilon_2$, as following:

$$I(\omega) = \sqrt{2}\omega \left[\sqrt{\varepsilon_1^2(\omega) + \varepsilon_2^2(\omega)} - \varepsilon_1(\omega)\right]^{1/2} \quad (3)$$

Which depends on $\varepsilon_1$ and $\varepsilon_2$ and thus on the energy. All other optical constants can also be obtained. The relations above are the theoretical basis of band structure and optical properties, which reflected the mechanism of luminescence spectral caused by electronic transition between various energy levels.

## Author Contributions

## Additional Information

Competing financial interests: The authors declare no competing financial interests.

Table 1. The distances from the $Ag_3PO_4(100)$ surface to pure and doped GR sheets after optimization and the formation energy $E_f$ (eV) for the $GR/Ag_3PO_4(100)$ and doped $GR/Ag_3PO_4(100)$ composites.

| Composites | $E_f$ (eV) | Interface distance (Å) | |
|---|---|---|---|
| | | d1 (Å) | d2 (Å) |
| $GR/Ag_3PO_4(100)$ | -0.77 | 2.61 | 2.65 |
| $B_C$-$GR/Ag_3PO_4(100)$ | -5.12 | 2.63 | 2.68 |
| $N_C$-$GR/Ag_3PO_4(100)$ | -5.48 | 3.12 | 3.15 |
| $S_C$-$GR/Ag_3PO_4(100)$ | 2.22 | 1.73 | 2.71 |
| $P_C+V_C$-$GR/Ag_3PO_4(100)$ | -7.34 | 2.63 | 2.68 |

d1 and d2 denote the farthest and closest distances from the $Ag_3PO_4(100)$ surface to pure/doped GR sheets, respectively.

Table 2. The band gap and Bader charge analysis for the undoped and doped GR/Ag$_3$PO$_4$(100) composites.

| Composites | $E_g$ (eV) | Bader charge (e) | |
|---|---|---|---|
| | | B/N/S/P+V$_C$-GR | Ag$_3$PO$_4$(100) |
| GR/Ag$_3$PO$_4$(100) | 0.53 | 0.39 | -0.39 |
| B$_C$-GR/Ag$_3$PO$_4$(100) | 0.12 | 0.15 | 0.15 |
| N$_C$-GR/Ag$_3$PO$_4$(100) | 1.44 | 0.61 | -0.61 |
| S$_C$-GR/Ag$_3$PO$_4$(100) | 0.38 | 0.47 | -0.47 |
| P$_C$+V$_C$-GR/Ag$_3$PO$_4$(100) | 0.12 | 1.48 | -1.48 |

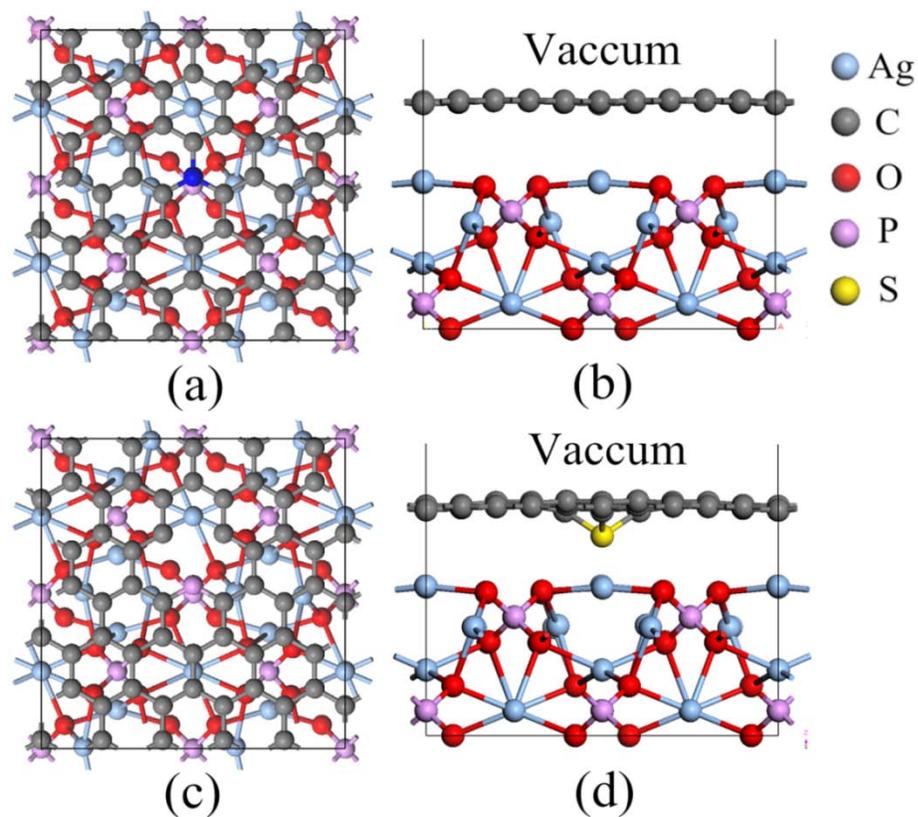

Fig. 1. Crystal structures of GR/Ag$_3$PO$_4$(100) composites. (a) Top view. The blue atom represents the location of dopant (B, N, S, and P). (b) Side view. (c) Top view of P$_C$+V$_C$-GR/Ag$_3$PO$_4$ (100) composite. One C atom bonded to P atom is removed to eliminate the large deformation of doped GR sheet. (d) Side view of crystal structure after optimization for S$_C$-GR/Ag$_3$PO$_4$ (100) composite. The S dopant leads to a local distinct deformation of the GR sheet.

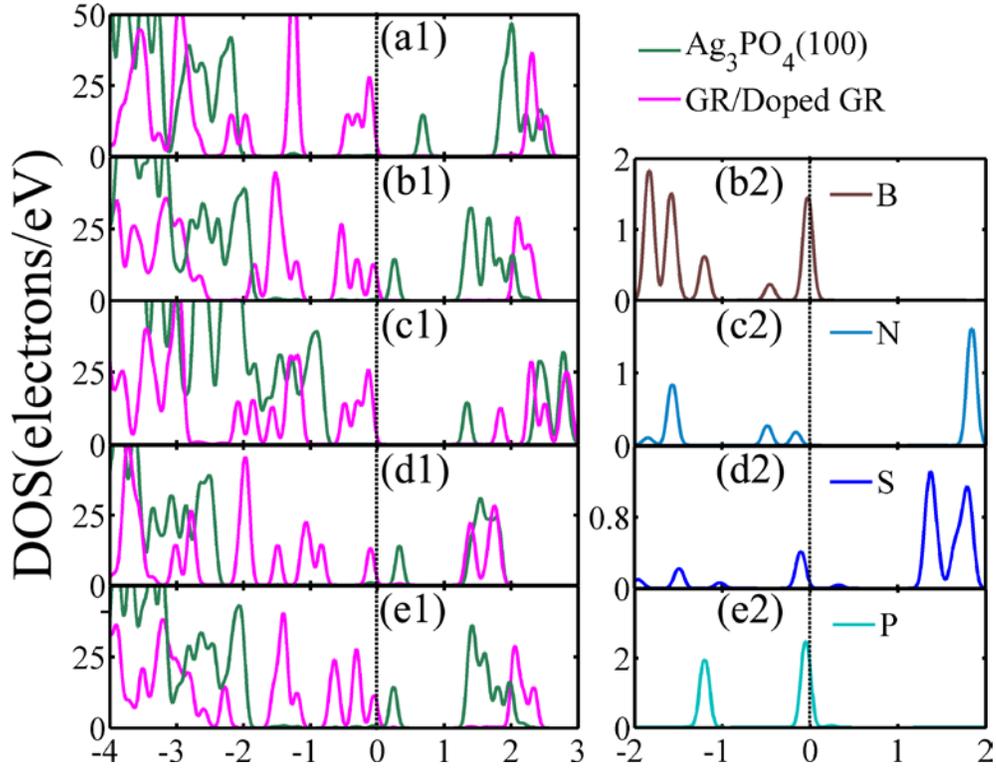

Fig. 2. Density of states (DOSs). Left: The project DOSs for $Ag_3PO_4(100)$ and pure/doped GR sheets. (a1)-(e1) are for GR/$Ag_3PO_4(100)$, $B_C$-, $N_C$-, $S_C$-, and $P_C+V_C$-GR/$Ag_3PO_4(100)$ composites, respectively. The magenta and green curves are for pure/doped GR sheets and $Ag_3PO_4(100)$ surface, respectively. It is clear that the near-gap electronic structures of the doped-GR/$Ag_3PO_4(100)$ composites are dependent on the dopants. Right: (b2)-(e2) DOSs of B, N, S, and P, respectively. The vertical dashed line indicates the Fermi level and the Fermi level is set at zero energy.

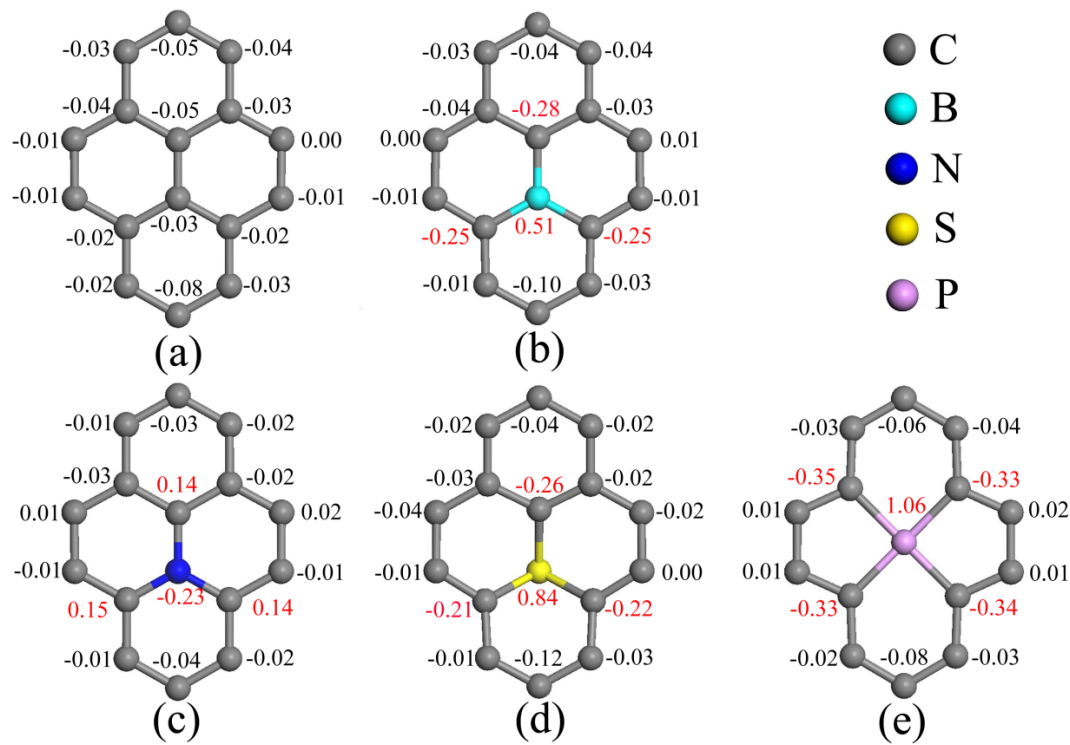

Fig. 3. The Mulliken population charges on the doped atoms and the atoms around the doped atom in GR sheet. (a)-(e) are for undoped, $B_C$-, $N_C$-, $S_C$-, and $P_C+V_C$-doped $GR/Ag_3PO_4(100)$ composites, respectively. The red digits denote the significant variations of Mulliken population charges, relative to the corresponding positions in the $GR/Ag_3PO_4(100)$ composite. It is clear that some C atoms become positively charged, while others are negative in these composites. Those atoms, especially the dopants, with positive effective charge will facilitate the adsorption of some species with negative charges, thus becoming active sites.

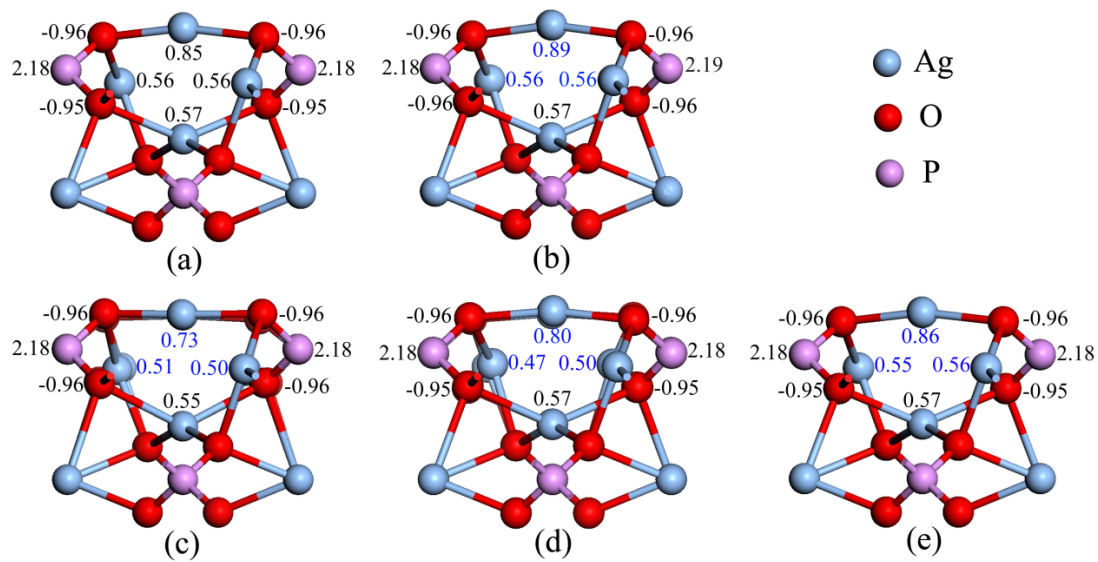

Fig. 4. The Mulliken population charges on Ag, O, and P atoms, which have changed in the $Ag_3PO_4(100)$ surface. (a)-(e) are for undoped, $B_C$-, $N_C$-, $S_C$-, and $P_C+V_C$-doped $GR/Ag_3PO_4(100)$ composites, respectively. The blue digits represent the changed Mulliken population charges on Ag atoms near the interface. The varistion of Mulliken population charges on P atoms shows that the dopants in GR sheets change the charge redistribution of the $Ag_3PO_4(100)$ surface.

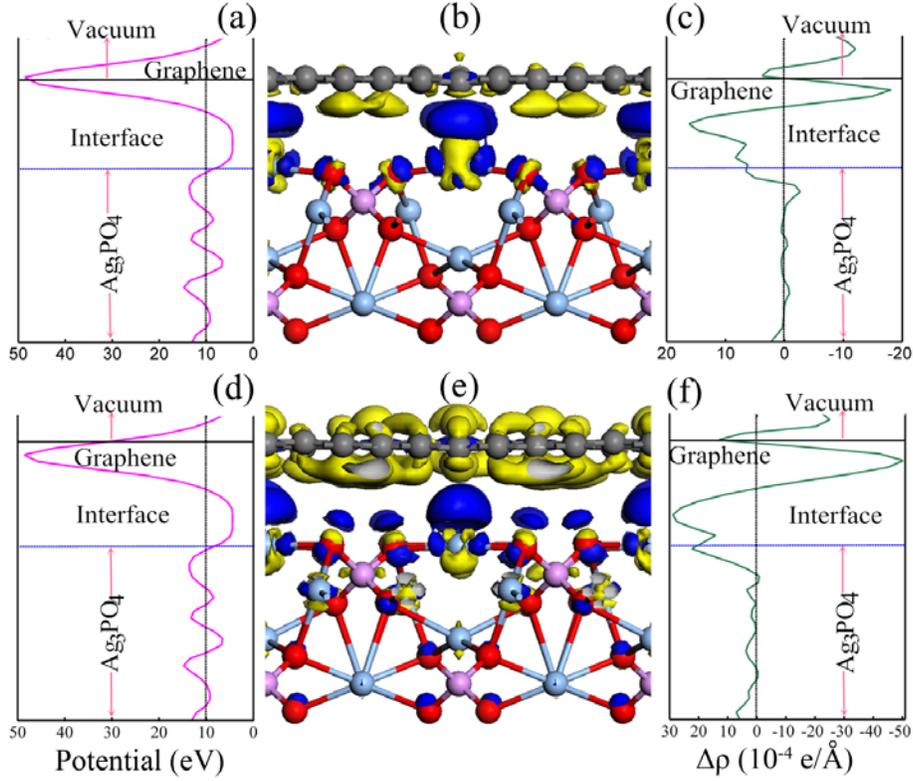

Fig. 5. Planar averaged self-consistent electrostatic potential as a function of position in the $Z$-direction: (a) and (d) are for $B_C$- and $N_C$-GR/$Ag_3PO_4$(100) composites, respectively. The higher potential near the doped GR sheets leads to a potential well formed at interfaces. 3D charge density difference with an isovalue of 0.006 e/Å$^3$: (b) and (e) are for $B_C$- and $N_C$-GR/$Ag_3PO_4$(100) composites, respectively. Blue and yellow is surfaces represent charge accumulation and depletion in the space. Planar averaged charge density difference as a function of position in the $Z$-direction: (c) and (f) are for $B_C$- and $N_C$-GR/$Ag_3PO_4$(100) composites, respectively. The charge difference shows that charge transfer occurs inversely from the $Ag_3PO_4$ surface to GR and induces the electrostatic potential inverse at the interface.

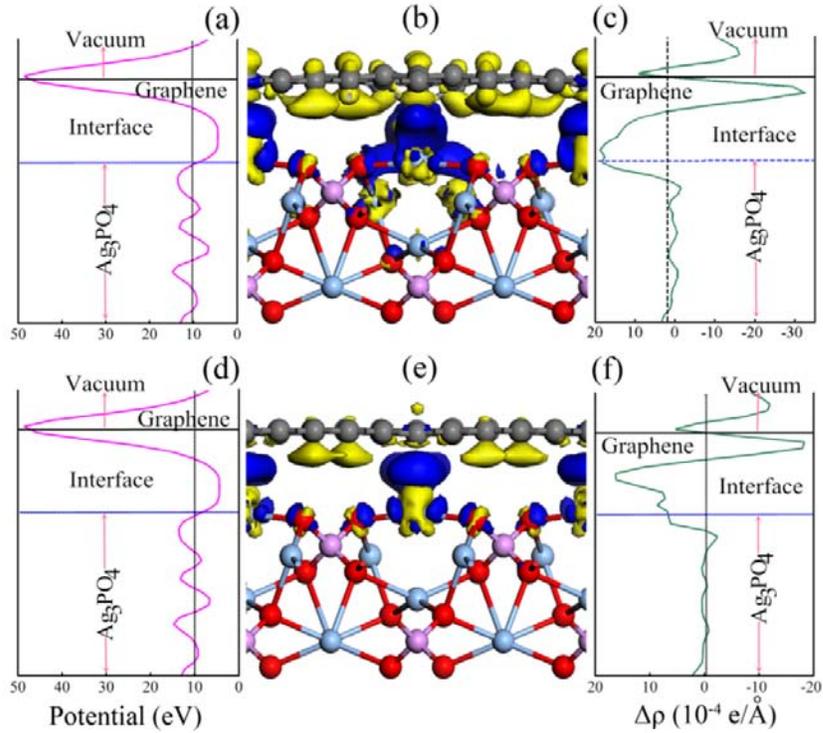

Fig. 6. Planar averaged self-consistent electrostatic potential as a function of position in the $Z$-direction: (a) and (d) are for $S_C$- and $P_C+V_C$-GR/Ag$_3$PO$_4$(100) composites, respectively. The higher potential near the doped GR sheets leads to a potential well formed at interfaces. 3D charge density difference with an isovalue of 0.006 e/Å$^3$: (b) and (e) are for $S_C$- and $P_C+V_C$-GR/Ag$_3$PO$_4$(100) composites, respectively. Blue and yellow is surfaces represent charge accumulation and depletion in the space. Planar averaged charge density difference as a function of position in the $Z$-direction: (c) and (f) are for $S_C$- and $P_C+V_C$-GR/Ag$_3$PO$_4$(100) composites, respectively. The charge difference shows that charge transfer occurs inversely from the Ag$_3$PO$_4$ surface to GR and induces the electrostatic potential inverse at the interface.

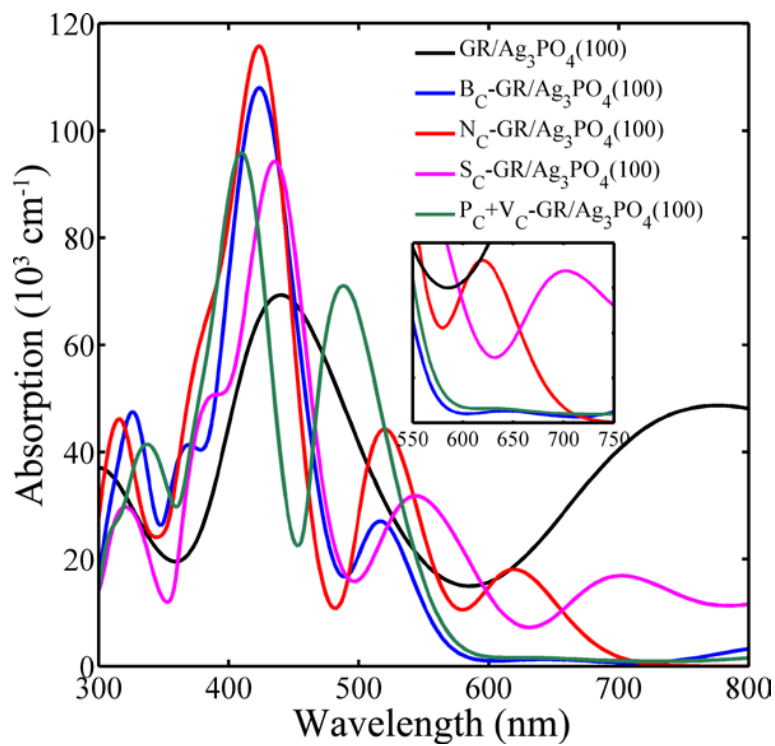

Fig. 7. Absorption spectra of the undoped, $B_C$-, $N_C$-, $S_C$-, and $P_C+V_C$-GR/$Ag_3PO_4$(100) composites for the polarization vector perpendicular to the surface. The inset of shows that the $N_C$- and $S_C$-GR/$Ag_3PO_4$(100) composites display strong absorption in the range 550 to 750 nm. The higher resonant-like peaks at about 400~440 nm, due to doping, are especially beneficial to enhance the photocatalytic performance of the doped GR/$Ag_3PO_4$(100) composites, according to the fact that the light with wavelength of ~440 nm might be the most appropriate visible light for generation of radical species in the GR/$Ag_3PO_4$(100) composite. The N atom may be most appropriate doping for the GR/$Ag_3PO_4$ photocatalyst.